\documentstyle[12pt,epsf]{article}
\renewcommand{\baselinestretch}{1.0}

\newcommand{\be}{\begin{equation}}
\newcommand{\ee}{\end{equation}}
\begin{document}
\topmargin 0pt
\oddsidemargin=-0.4truecm
\evensidemargin=-0.4truecm
\renewcommand{\thefootnote}{\fnsymbol{footnote}}

\newpage
\setcounter{page}{1}
\begin{titlepage}     
\vspace*{-2.0cm}
\begin{flushright}
\vspace*{-0.2cm}
\end{flushright}
\vspace*{0.5cm}

\begin{center}
{\Large \bf LMA MSW solution of the solar neutrino problem and first 
KamLAND results}
\vspace{0.5cm}

{P. C. de Holanda$^{1}$ and  A. Yu. Smirnov$^{2,3}$\\
\vspace*{0.2cm}
{\em (1) Instituto de F\'\i sica Gleb Wataghin - UNICAMP, 
13083-970 Campinas SP, Brasil}\\
{\em (2) The Abdus Salam International Centre for Theoretical Physics,  
I-34100 Trieste, Italy }\\
{\em (3) Institute for Nuclear Research of Russian Academy 
of Sciences, Moscow 117312, Russia}

}
\end{center}

\begin{abstract}
The first KamLAND results are in a very good agreement with the
predictions made on the basis of the solar neutrino data and the LMA
realization of the MSW mechanism. We perform a combined analysis of
the KamLAND (rate, spectrum) and the solar neutrino data with a free
boron neutrino flux $f_B$.  The best fit values of neutrino parameters
are $\Delta m^2 = 7.3 \cdot 10^{-5}$ eV$^2$, $\tan^2 \theta = 0.41$ and
$f_B = 1.05$ with the $1\sigma$ intervals: $\Delta m^2 = (6.2 - 8.4)
\cdot 10^{-5}$ eV$^2$, $\tan^2 \theta = 0.33 - 0.54$.  We find the
$3\sigma$ upper bounds: $\Delta m^2 < 2.8 \cdot 10^{-4}$ eV$^2$ and
$\tan^2 \theta < 0.84$, and the lower bound $\Delta m^2 > 4 \cdot
10^{-5}$ eV$^2$.  At 99\% C.L. the KamLAND spectral result splits the
LMA region into two parts with the preferred one at $\Delta m^2 <
10^{-4}$ eV$^2$.  The higher $\Delta m^2$ region is accepted at about
$2\sigma$ level.  We show that effects of non-zero 13-mixing, $\sin^2
\theta_{13} \leq 0.04$, are small leading to a slight improvement of the
fit in higher $\Delta m^2$ region.  In the best fit point we predict
for SNO: CC/NC $= 0.33 ^{+0.05}_{-0.03}$ and $A_{DN}^{SNO} = 2.8 \pm
0.8\%$ (68\% C.L.), and $A_{DN}^{SNO} < 9 \%$ at the $3\sigma$
level. Further improvements in the determination of the oscillation 
parameters are discussed and implications 
of the solar neutrino and KamLAND results are considered.

\end{abstract}

%%%%%%%%%%%%%%%%%%%%%%%%%%%%%%%%%%%%%%%%%%%%%%%%%%%%%%%%%%%%%%%%%%%
\end{titlepage}
\renewcommand{\thefootnote}{\arabic{footnote}}
\setcounter{footnote}{0}
\renewcommand{\baselinestretch}{0.9}
%%%%%%%%%%%%%%%%%%%%%%%%%%%%%%%%%%%%%%%%%%%%%%%%%%%%%%%%%%%%%%%%%%%

%%%%%%%%%%%%%%%%%%%%%%%%%%%%%%%%%%%%%%%%%%%%%%%%%%%%%%%%%%%%%%%%%%
\section{Introduction} 
%%%%%%%%%%%%%%%%%%%%%%%%%%%%%%%%%%%%%%%%%%%%%%%%%%%%%%%%%%%%%%%%%%%

The first KamLAND results~\cite{kamf} are the last (or almost last)
step in resolution of the long-standing solar neutrino
problem~\cite{john}.  ``In the context of two flavor neutrino
oscillations with the CPT invariance, the KamLAND results exclude all
oscillation solutions of this problem but the `Large mixing angle'
solution"~\cite{kamf}.  In fact, KamLAND excludes also all {\it
non-oscillation} solutions based on neutrino spin-flip in the magnetic
fields of the Sun, on the non-standard neutrino interactions, etc..
More precisely, KamLAND excludes them  as the dominant
mechanisms of the solar neutrino conversion.

Soon after the suggestion of the MSW mechanism \cite{msw}, the ``MSW
triangle"~\cite{triangle} had been constructed in the $\Delta m^2 -
\sin^2 2\theta$ plane which corresponds to 1/4 - 1/3 suppression of
the Argon production rate~\cite{Cl}.  The region around the upper
right corner of this triangle, or large $\sin^2 2\theta$ part of its
horizontal side (``the adiabatic solution") is what we call now the LMA
region.

Kamiokande~\cite{kamioka}, SAGE~\cite{sage} and GALLEX~\cite{gno} have
disintegrated the triangle into pieces excluding some parts of
oscillation parameter space.  First, the results from Kamiokande (the
absence of a strong spectrum distortion as well as the absence of
strong Day-Night effect) have chipped off the LMA region. Then Gallium
experiments SAGE and GALLEX, confirming LMA, have splitted the diagonal
side of the triangle into SMA and LOW.  Apart from that vacuum
oscillation solutions (VO) were always present.

For a long time SMA was the favored solution and LMA was considered as
an non excluded possibility. Things changed in 1998 when
Super-Kamiokande~\cite{SK} has testified against SMA (the flatness of
spectrum and the absence of a peak in the zenith angle distribution of
events in the Earth core bin).  Soon after that, the analysis of the
Super-Kamiokande data for 708 days of operation allowed us~\cite{bks}
to conclude that solar neutrino results provide hints that the LMA
solution could be correct (see also analysis in~\cite{valle}).  Since
that time LMA continuously reinforced its position.

Super-Kamiokande~\cite{SK} and then SNO~\cite{sno} have accomplished
genesis of the triangle. They have practically excluded SMA,
disfavored LOW and VO, and shifted the LMA region to larger $\Delta
m^2$ (small Day-Night asymmetry at Super-Kamiokande) and smaller
mixings (small ratio of CC/NC events at SNO). \\

By the time of the KamLAND announcement, the solar neutrino
data~\cite{Cl,sage,gno,SK,sno,sno-nc,sno-dn} have definitely selected
LMA as the most favorable solution based on neutrino mass and
mixing~\cite{mswlma,us:msw}. The best fit point from the free boron neutrino
flux fit~\cite{us:msw} is
\be
\Delta m^2 = 6.15 \cdot 10^{-5} {\rm eV}^2 ,
~~~ \tan^2 \theta = 0.41,~~~ f_B = 1.05, 
\label{bfsol}
\ee
where $f_B \equiv F_B/F_B^{SSM}$ is the boron neutrino flux in the
units of the Standard Solar Model predicted flux~\cite{ssm}.

On basis of the solar neutrino results (and the assumption of the CPT
invariance) predictions for the KamLAND experiment have been
calculated. A significant suppression of the signal was expected in
the case of the LMA solution.  The predicted ratio of
the numbers of events with the visible (prompt) energies, $E_p$,  
above 2.6 MeV  with and without oscillations equals~\cite{us:msw}:
\be
R_{KL}^{LMA} = 0.65 ^{+ 0.09}_{-0.40} ~~~(3\sigma).
\label{klr}
\ee
For other solutions of the solar neutrino problem one expected $R_{KL} = 0.9 -
1$, where the deviation from 1 can be due to the effect of nonzero
1-3 mixing.

In the best fit point (\ref{bfsol}) the predicted spectrum has (i) a peak
at $E_{p} \approx (3.0 - 3.6)$ MeV, (ii) a suppression of the number of
events near the threshold energy $E_{p} \approx 2.6$ MeV and (iii) 
a significant suppression of the signal (with respect to the
no-oscillation case) at the high energies: $E_{p} > (4 - 5)$ MeV
\cite{us:msw}.  No distortion of the spectrum is expected for the
other solutions.\\

The first KamLAND results: both the total number of events and
spectrum shape~\cite{kamf}, are in a very good agreement with
predictions:
\be
R_{KL}^{exp} = 0.611 \pm 0.094~.
\label{bestkamrates}
\ee
The spectral data (although not yet precise) reproduce well the
features described above.  As a result, the allowed ``island" in the
$\Delta m^2 - \sin^2 2\theta$ plane with the best fit KamLAND point
covers the best fit point from the solar neutrino
analysis~\cite{kamf}.\\

In this paper we present our analysis of the KamLAND data as well as
combined analysis of the solar neutrinos  and KamLAND results.  The impact of
KamLAND on the LMA MSW solution is studied.  We consider implications
of the combined (solar neutrinos + KamLAND) analysis and make
predictions for the future measurements.

During preparation of this paper, several studies of the first KamLAND
results and solar neutrino data have been published~\cite{after} -
\cite{alia}. Our conclusions are in a good agreement with results of
those papers, although there are some differences. Detailed comparison
of results will be done later.

%%%%%%%%%%%%%%%%%%%%%%%%%%%%%%%%%%%%%%%%%%%%%%%%%%%%%%%%%%%%%%%%
\section{KamLAND}
%%%%%%%%%%%%%%%%%%%%%%%%%%%%%%%%%%%%%%%%%%%%%%%%%%%%%%%%%%%%%%%%

In a given energy bin $a$ $(a = 1 , ....  13)$ 
the  signal at KamLAND  is determined  by   
\be
N_a = A \sum_i  \int_{E_a}^{E_a + \Delta E} dE_p \int dE_p' 
P_i  F_i \sigma f(E_p, E_p')~,  
\label{n-rate}
\ee
where $\Delta E = 0.425$ MeV, 
\be
P_i  =  \left(1 - \sin^2 2\theta_{12} \sin^2  
\frac{\Delta m^2_{12} L_i}{4E} \right) 
\label{survpr}
\ee
is the vacuum oscillation survival probability for $i$ reactor
situated at the distance $L_i$ from KamLAND, $F_i$ is the flux from
$i$ reactor, $\sigma$ is the cross-section of $\bar \nu p \rightarrow
e^+ n$ reaction, $E_p$ is the observed prompt energy, $E_p'$ is the
true prompt energy, $f(E_p, E_p')$ is the energy resolution function,
$A$ is the factor which takes into account the fiducial volume, the
time of observation, etc..  We sum over all reactors contributing
appreciable to the flux at KamLAND.

The suppression factor of the total number of (reactor neutrino)
events above certain threshold is defined as:
\be
R_{KL}(\Delta m^2, \tan^2 \theta) \equiv \frac{N(\Delta m^2, \tan^2
\theta)}{N_0}~,
\ee
where $N = \sum_a N_a$, $N_a$ is given in (\ref{n-rate}) and $N_0$ is
the total number of events in the absence of oscillations (following KamLAND
we will call $N$ and $N_0$ the rates).\\

%%%%%%%%%%%%%%%%%%%%%%%%%%%%%%%%%%%%%%%%%%%%%%%%%%%%%%%%%%%%%%%%%%%%%%%%%%%%%
\noindent
{\it 1).  KamLAND spectrum. High threshold.}  We perform $\chi^2$
analysis of the KamLAND spectrum, defining
\begin{equation}
\chi^2_{spec}=\sum_{a=1...13}\sum_{b=1...13} (N_a - N_a^{th}) 
\sigma^{-2}_{ab}(N_b - N_b^{th})~,   
\label{chisp}
\end{equation}
where $\sigma_{ab}$ is the covariance matrix in which the systematic
uncertainties were propagated to the spectral bins.
 
We find that for $E_p \geq  2.6$ MeV the minimum of $\chi^2_{spec}$ is
achieved for
\be
\Delta m^2 = 7.34 \cdot 10^{-5}  {\rm eV}^2 , ~~~  \tan^2 \theta = 0.453,    
\label{hth1}
\ee 
and in this point $\chi^2/ d.o.f. = 2.92/11$. Notice that in contrast
with the KamLAND result~\cite{kamf} our best fit mixing deviates from
the maximal mixing.

\begin{figure}[ht]
\centering\leavevmode
\epsfxsize=.8\hsize
\epsfbox{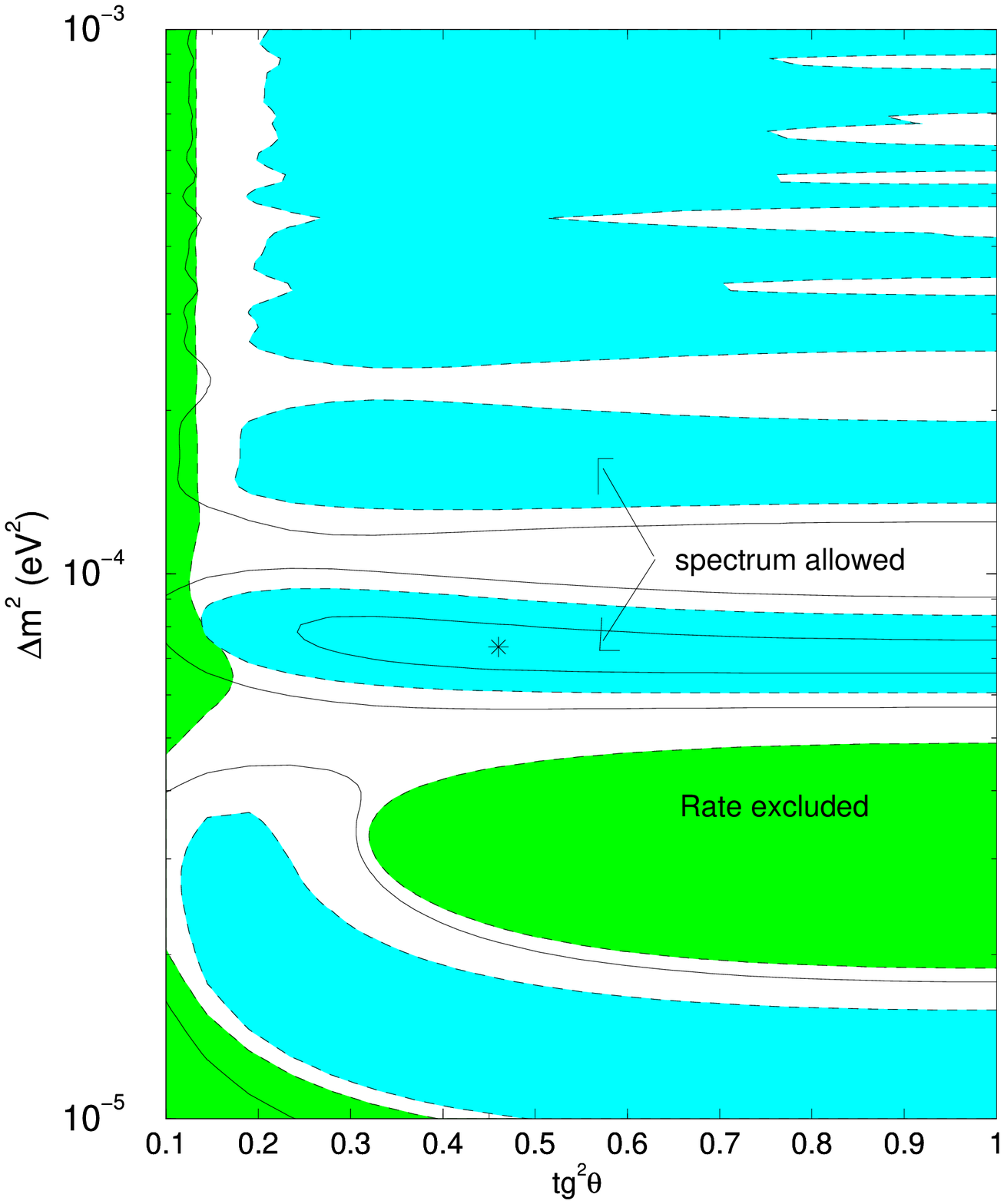}
\caption{The KamLAND spectrum analysis for $E_p > 2.6$ MeV. 
Shown are the allowed regions of oscillation parameters at 68\% (inner
solid lines), 95\% (grey) and 99\% C.L.  (outer solid lines).  The
best fit point is indicated by star.  Also shown are the regions
excluded by the rate analysis at $95\%$ C.L. (dark).}
\label{fig:chi2kaml}
\end{figure}

We present in fig.~\ref{fig:chi2kaml} the contours of constant
confidence level with respect to the best fit point (\ref{hth1}) in
the $(\Delta m^2 - \tan^2 \theta)$ plane using relation: $\chi^2 =
\chi^2_{min} + \Delta \chi^2$, where $\Delta \chi^2 = 1, 3.84$ and  $6.63$
for $1\sigma$, $95\%$ and $99\%$ C.L. correspondingly.

The contours manifest an oscillatory pattern in $\Delta m^2$ in spite
of a strong averaging effect which originates from large spread in
distances from different reactors. The pattern can be described in
terms of oscillations with certain effective distance, $L_{eff}$, and effective
oscillation phase $\phi_{eff}$: 
\be
\phi_{eff} = \frac{\Delta m^2_{12} L_{eff}}{4E}, ~~~L_{eff} \approx 165~ {\rm km}.  
\label{effph}
\ee
$L_{eff}$ corresponds to the distance  
between KamLAND and the closest set of reactors which provides 
the large fraction  of  the 
antineutrino flux. Notice that $L_{eff}$  is smaller than  the average (weighted with 
power) 
distance 189 km.\\ 

Let us consider the 95\% allowed regions.
 
(i) The lowest ``island" allowed by KamLAND with  $\Delta m^2 < 2 \cdot
10^{-5}$ eV$^2$  corresponds to the oscillation phase $\phi_{eff} <
\pi/2$. This region is excluded by the absence of significant
day-night asymmetry of the Super-Kamiokande signal. In this domain,
the predicted  asymmetry at SNO, $A_{DN}^{SNO} > 17\%$, is still
consistent with data.

(ii) The second allowed region, $\Delta m^2 = (5 - 10) \cdot 10^{-5}$
eV$^2$, corresponds to the first oscillation maximum, $\phi_{eff} \sim
\pi$ (maximum of the survival probability). It contains the best fit point.

(iii) The third island is at $\Delta m^2 = (13 - 23) \cdot 10^{-5}$
eV$^2$: it corresponds to the oscillation maximum (second maximum 
of the survival probability) with $\phi_{eff} \sim 2 \pi$.

There is a continuum of the allowed regions above $\Delta m^2 \sim 3
\cdot 10^{-4}$ eV$^2$.  The third region merges with the continuum
at 99\% CL.

At the $1\sigma$ level the second island is the only allowed
region. \\

%%%%%%%%%%%%%%%%%%%%%%%%%%%%%%%%%%%%%%%%%%%%%%%%%%%%%%%%%%%%%%%%%%%%%%%%%%%
\noindent
{\it 2) KamLAND rate.} In fig.\ref{fig:chi2kaml} we show the regions
excluded by the KamLAND rate at $95 \%$ C.L..  The borders of
these regions coincide with contours of constant $R_{KL}^{max} = 0.80$
and $R_{KL}^{min} = 0.42$ obtained in \cite{us:msw}).  The exclusion
region at $\Delta m^2 \sim (2- 5) \cdot 10^{-5}$ eV$^2$ corresponds to
the first oscillation minimum (minimum of the survival probability) 
at KamLAND ($\phi_{eff} \sim \pi/2$).
Here the suppression of the signal is too strong.  Another region of
significant suppression (second oscillation minimum with $\phi_{eff} =
3\pi/2$) is at $\Delta m^2 \sim (9 - 12) \cdot 10^{-5}$ eV$^2$. \\

\begin{figure}[ht]
\centering\leavevmode
\epsfxsize=.8\hsize
\epsfbox{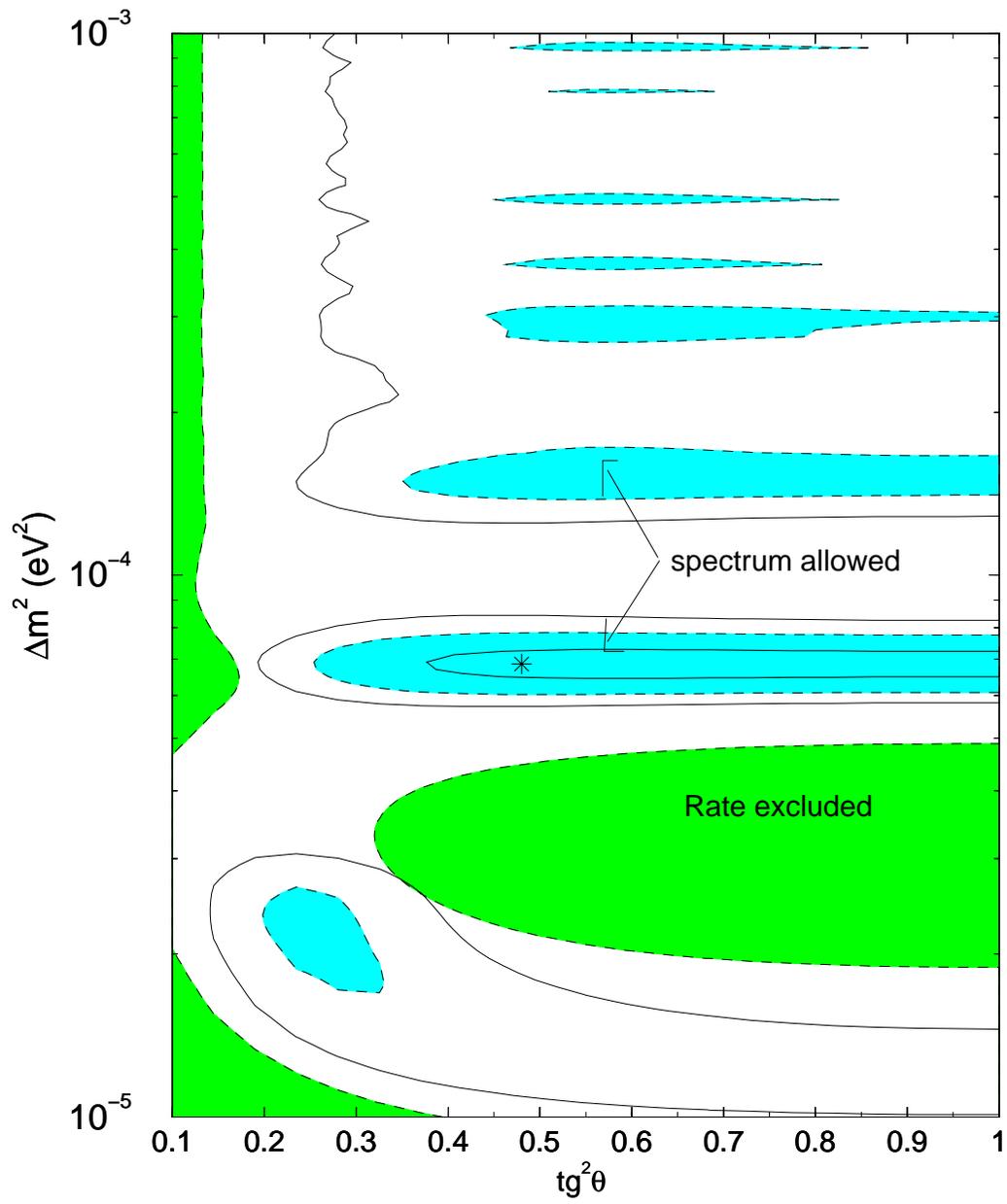}
\caption{The same as in fig. 1 for $E_p > 0.9$ MeV.}
\label{fig:chi2kaml_lt}
\end{figure}

%%%%%%%%%%%%%%%%%%%%%%%%%%%%%%%%%%%%%%%%%%%%%%%%%%%%%%%%%%%%%%%%%%%%%%%%%%%
\noindent
{\it 3).  KamLAND spectrum. Low threshold.}  We analyze the spectrum
with the threshold $E_p^{th} = 0.9$ MeV, using the same prescription for
the contribution of the geological neutrinos as in \cite{kamf}.  The
best fit point shifts with respect to (\ref{hth1}) to larger mixing
and smaller mass squared difference:
\be
\Delta m^2 = 6.86 \cdot 10^{-5}  {\rm eV}^2, ~~~  \tan^2 \theta = 0.48 \,. 
\label{hth2}
\ee 
In fig.~\ref{fig:chi2kaml_lt} we show the allowed regions at 68 \%,
95\% and 99\% confidence levels.

With lowering the threshold the sensitivity to the spectrum shape
increases: the measurements become sensitive not only to the dominant
oscillation peak but also to the lower energy oscillation maximum.  As a
result, the analysis leads to stronger exclusion of the oscillation
parameter space.  In particular, maximal and large mixing parts become
less favored than in the case of high threshold. \\

\noindent
{\it 4).  KamLAND spectrum and rate.}  Following procedure in
\cite{kamf} we have performed also combined analysis of spectrum and
rate introducing the free normalization parameter of the spectrum,
$R_{KL}$, and defining the $\chi^2$ as
\be
\chi^2_{spec,R} = \chi^2_{spec} + \chi^2_{R},
\label{s-klr}
\ee
where
\be
\chi^2_{R} = \left(\frac{R_{KL} - 0.611}{0.094}\right)^2 .
\label{s-klr2}
\ee
We find results which are very close to those from our spectrum
analysis.  In particular, the best fit value of mixing is $\tan^2
\theta = 0.48$ and $\Delta m^2 = 7.31\cdot 10^{-5}$ eV$^2$.\\

As it follows from our consideration here, the values of oscillation
parameters extracted from the KamLAND data, $(\Delta m^2, ~~ \tan^2
\theta)_{KL}$, are in a very good agreement with the values  from
independent solar neutrino analysis $(\Delta m^2, ~~ \tan^2
\theta)_{sun}$, For the best fit points (\ref{bfsol}), (\ref{hth1})
(\ref{hth2}) we conclude that within $1\sigma$ (see 
fig.~\ref{fig:chi2kaml} and  fig.~\ref{fig:chi2kaml_lt}) 
\be
(\Delta m^2, ~~ \tan^2 \theta)_{KL} = (\Delta m^2, ~~ \tan^2
\theta)_{sun}.
\label{equality}
\ee
At the same time, the data do not exclude that the solar and KamLAND
parameters are  different, and moreover, the difference still can be
large.  For instance,  $(\Delta m^2, ~~ \tan^2 \theta)_{KL}$ can
coincide with the present best fit point or be  in the high 
$\Delta m^2$ island, whereas $(\Delta m^2, ~~ \tan^2 \theta)_{sun}$ can be  
at  lower $\Delta m^2$.

%%%%%%%%%%%%%%%%%%%%%%%%%%%%%%%%%%%%%%%%%%%%%%%%%%%%%%%%%%%%%%%%
\section{Solar Neutrinos}
%%%%%%%%%%%%%%%%%%%%%%%%%%%%%%%%%%%%%%%%%%%%%%%%%%%%%%%%%%%%%%%%

We use the same data set and the same procedure of analysis as in our
previous publication~\cite{us:msw}.  Here the main ingredients of the
analysis are summarized.

The data sample consists of

\noindent
- 3 total rates: (i) the $Ar$-production rate, $Q_{Ar}$, from
Homestake~\cite{Cl}, (ii) the $Ge-$production rate, $Q_{Ge}$ from SAGE
\cite{sage} and (iii) the combined $Ge-$production rate from GALLEX
and GNO \cite{gno};

\noindent
- 44 data points from the zenith-spectra measured by Super-Kamiokande
during 1496 days of operation \cite{SK};

\noindent
- 38 day-night spectral points from SNO \cite{sno-nc,sno-dn}.

Altogether the solar neutrino experiments provide us with 81 data points.

All the solar neutrino fluxes, but the boron neutrino flux, are taken
according to SSM BP2000 \cite{ssm}.  The boron neutrino flux is
treated  as a free parameter.  For the $hep-$neutrino flux we take
fixed value $F_{hep} = 9.3 \times 10^{3}$ cm$^{-2}$ s$^{-1}$
\cite{ssm,hepfl} .

Thus, in our analysis of the solar neutrino data as well as in the
combined analysis of the solar and KamLAND results we have three fit
parameters: $\Delta m^2$, $\tan^2\theta$ and $f_B$.\\

We define the contribution of the solar neutrino data to $\chi^2$ as
\be
\chi^2_{sun} = \chi^2_{rate} +   \chi^2_{SK} + \chi^2_{SNO}, 
\label{chi-def}
\ee
where $\chi^2_{rate}$, $\chi^2_{SK}$ and $\chi^2_{SNO}$ are the
contributions from the total rates, the Super-Kamiokande zenith
spectra and the SNO day and night spectra correspondingly.  The main
result of analysis performed in \cite{us:msw} is given here in
Eq. (\ref{bfsol}).

%%%%%%%%%%%%%%%%%%%%%%%%%%%%%%%%%%%%%%%%%%%%%%%%%%%%%%%%%%%%%%%%%%%%%%%%%%%
\section{Solar neutrinos and KamLAND
\label{solakam}}
%%%%%%%%%%%%%%%%%%%%%%%%%%%%%%%%%%%%%%%%%%%%%%%%%%%%%%%%%%%%%%%%%%%%%%%%%%%

We have performed two different combined fits of the data from the
solar neutrino experiments and KamLAND. \\

\noindent
{\it 1) KamLAND rate and solar neutrino data.}  There are 81 (solar) +
1 (KamLAND) data points - 3 free parameters = 79 d.o.f.. We define
the global $\chi^2$ for this case as
\be
\chi^2_{sun,R} = \chi^2_{sun} + \chi^2_{R}, 
\label{s-klr3}
\ee
where $\chi^2_{sun}$ and $\chi^2_{R}$ are given in (\ref{chi-def})
and (\ref{s-klr2}). 
The minimum $\chi^2_{sun,R}(min)/d.o.f.  = 65.2/79$ corresponds to the
C.L. = 86.7\% . It appears at
\be
\Delta m^2 = 6.03 \cdot 10^{-5}  {\rm eV}^2, ~~~ \tan^2 \theta =
0.411, ~~~ f_B = 1.05 .
\label{hth3}
\ee 
This point  practically  coincides 
with what we have obtained from the solar
neutrino analysis only.

\begin{figure}[ht]
\centering\leavevmode
\epsfxsize=.8\hsize
\epsfbox{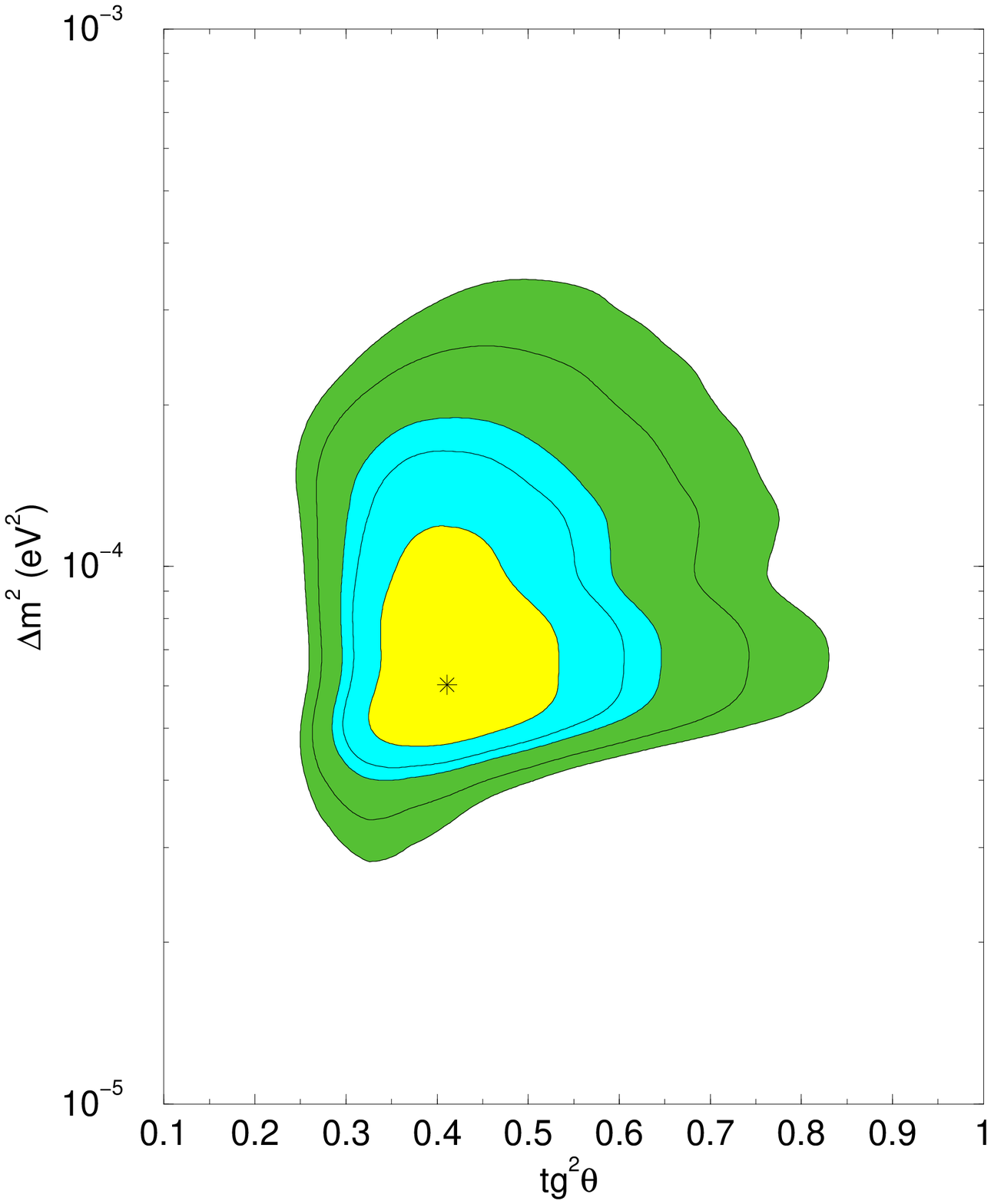}
\caption{The allowed regions in $\tan^2\theta - \Delta m^2$ 
plane from a combined analysis of the solar neutrino data and the
KamLAND rate, at 1$\sigma$, 90\%, 95\%, 99\% and 3$\sigma$ C.L..  The
best fit point is marked by star.
% is at $\tan^2 \theta = 0.41$ and 
%$\Delta m^2 = 6.05 \cdot 10^{-5} {\rm eV}^2$.
 }
\label{fig:chi2combrates}
\end{figure}

We construct the contours of constant confidence level in the $(\Delta
m^2 - \tan^2 \theta)$ plot using the following procedure.  We perform
minimization of $\chi^2_{sun,R}$ with respect to $f_B$ for each point
of the oscillation plane, thus getting $\chi^2_{sun,R}(\Delta m^2,
\tan^2 \theta)$.  Then the contours are defined by the condition
$\chi^2_{sun,R}(\Delta m^2, \tan^2 \theta) = \chi^2_{sun,R}(min) +
\Delta \chi^2$, where $\Delta \chi^2 = 2.3,\,4.61,\,5.99,\,9.21$ and
$11.83$ are taken for $1\sigma,\,90\%,\,95\%$ and $99\%$ C.L.  and
$3\sigma$.  The results are shown in fig.~\ref{fig:chi2combrates}.

According to the figure, the main impact of the KamLAND rate is
strengthening of bound on the allowed region from below due to strong
suppression of the KamLAND rate at $\Delta m^2 = (2 - 5) \cdot
10^{-5}$ eV$^2$ (see fig. \ref{fig:chi2kaml}) - region of the first
oscillation minimum in KamLAND.  The lines of constant confidence
level are shifted to larger $\Delta m^2$.  The KamLAND rate leads to
a distortion (shift to smaller mixings) of contours at $\Delta m^2 \sim
10^{-4}$ eV$^2$ where the second oscillation minimum at  KamLAND is
situated.  The upper part of the allowed region is modified rather
weakly.\\

%%%%%%%%%%%%%%%%%%%%%%%%%%%%%%%%%%%%%%%%%%%%%%%%%%%%%%%%%%%%%%%%%%%%%%%
\noindent
{\it 2).  KamLAND spectrum and solar neutrino data.}  We calculate
\be
\chi^2_{global} = \chi^2_{sun} +   \chi^2_{spec} ~, 
\label{chi-gl}
\ee
where $\chi^2_{spec}$ has been defined in (\ref{chisp}).  In this case
we have 81 (solar) + 13 (KamLAND) data points - 3 free parameters = 91
d.o.f..  The absolute minimum, $\chi^2_{global}(min) = 68.2$ (which
corresponds to a very high confidence level: $96.3\%$), is at
\be
\Delta m^2 = 7.32 \cdot 10^{-5}  {\rm eV}^2,~~~\tan^2 \theta = 0.409,
~~~f_B = 1.05.
\label{hth4}
\ee 
The best fit value of $\Delta m^2$ is slightly higher than that from the
solar data analysis. The solar neutrino data have higher sensitivity
to mixing,  whereas the KamLAND is more sensitive to $\Delta m^2$, as a
result, in (\ref{hth4}) the value of $\Delta m^2$ is close to the one
determined from the KamLAND data only (\ref{hth1}), whereas $\tan^2
\theta$ coincides with mixing determined from the solar neutrino results
(\ref{bfsol}).

\begin{figure}[ht]
\centering\leavevmode
\epsfxsize=.8\hsize
\epsfbox{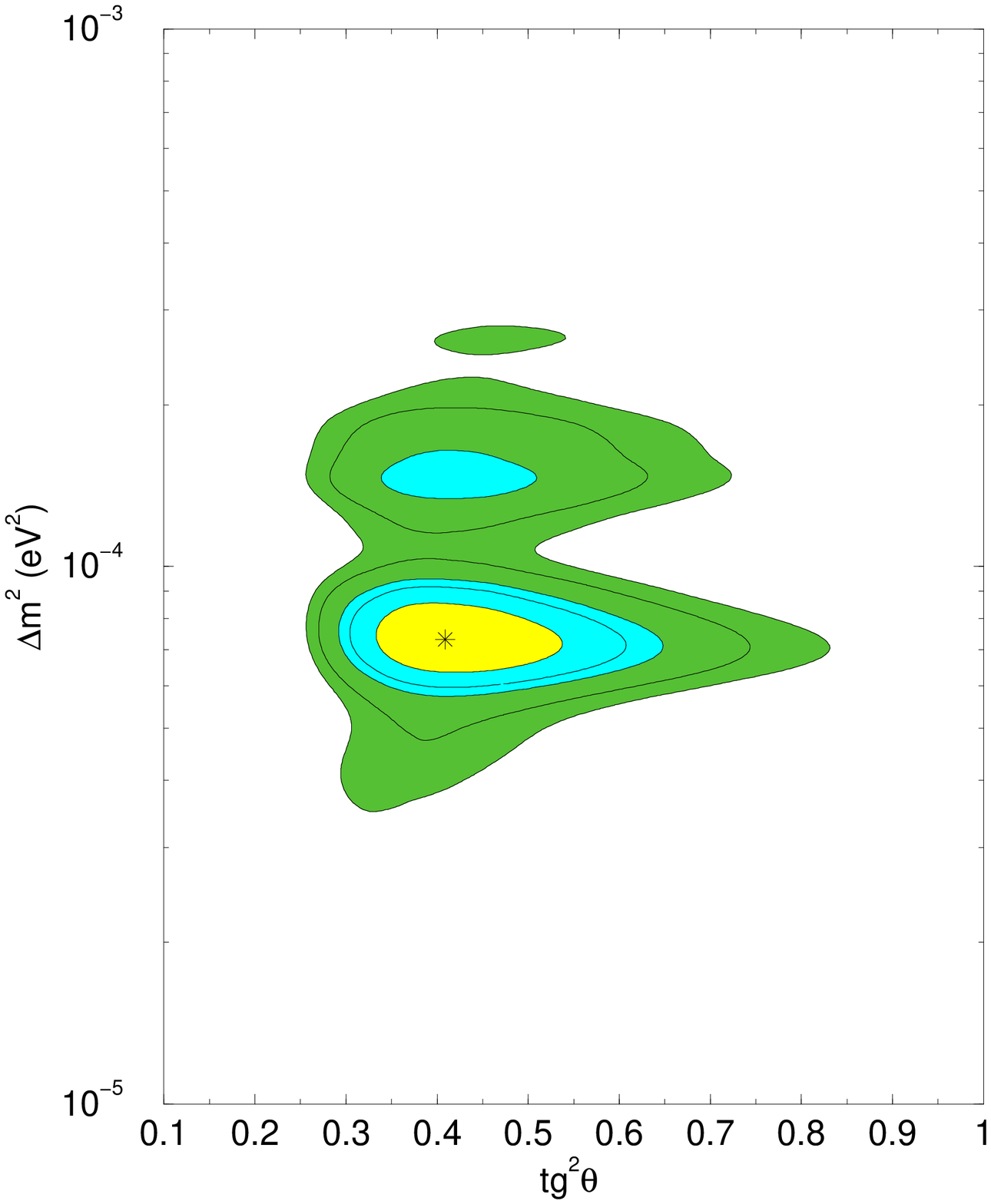}
\caption{The allowed regions in $\tan^2\theta - \Delta m^2$ 
plane, from a combined analysis of the solar neutrino data and the
KamLAND spectrum at 1$\sigma$, 90\%, 95\%, 99\% and 3$\sigma$ C.L..
The best fit point is marked by star.  }
\label{fig:chi2combspec}
\end{figure}

We construct the contours of constant confidence levels in the
oscillation plane, similarly to what we did for the fit of the solar
data and the KamLAND rate (fig. \ref{fig:chi2combspec}).

As compared with the solar data analysis, KamLAND practically has not
changed the upper bound on mixing, but strengthened the bound on
$\Delta m^2$.  At the $3\sigma$ level we get:
\be
\Delta m^2 < 2.8 \cdot 10^{-4}  {\rm eV}^2 , 
~~~ \tan^2 \theta < 0.84,~~~ 99.73\%~ {\rm C.L.}~.
\label{3sigm}
\ee  

The spectral data disintegrate the LMA region.  At the $3\sigma$ level
only a small spot is left in the range $\Delta m^2 > 2.5 \cdot
10^{-4}$ eV$^2$.

At the 99\% C.L. the rest of the region splits into two ``islands" which
we will refer to as the lower (l-) and higher (h-) LMA regions.
(Existence of these two islands can be seen already from an overlap of the
solar and KamLAND allowed regions in \cite{kamf}).  The l-region is
characterized by
\be
\Delta m^2 = (4.7  - 10) \cdot 10^{-5}  {\rm eV}^2 , 
~~~ \tan^2 \theta = 0.27 - 
0.75.
\label{lower}
\ee  
It contains the best fit point (\ref{hth4}).  The h-region is
determined by
\be
\Delta m^2 = (12 - 20) \cdot 10^{-5}  {\rm eV}^2 , 
~~~ \tan^2 \theta = 0.29 - 0.63,  
\label{higher}
\ee  
with best fit point at
\be
\Delta m^2 = 14.5 \cdot 10^{-5}  {\rm eV}^2 , ~~~ \tan^2 \theta = 0.41.   
\label{hth5}
\ee  
This point is accepted with respect to the global minimum (\ref{hth4})
at about $2\sigma$ level.

The excluded region between two islands corresponds to the second
oscillation minimum ($\phi_{eff} \sim 3\pi/2$) at $E_p \sim (3 - 4)
MeV$ which contradicts the spectral data. \\

\begin{figure}[ht]
\centering\leavevmode
\epsfxsize=.8\hsize
\epsfbox{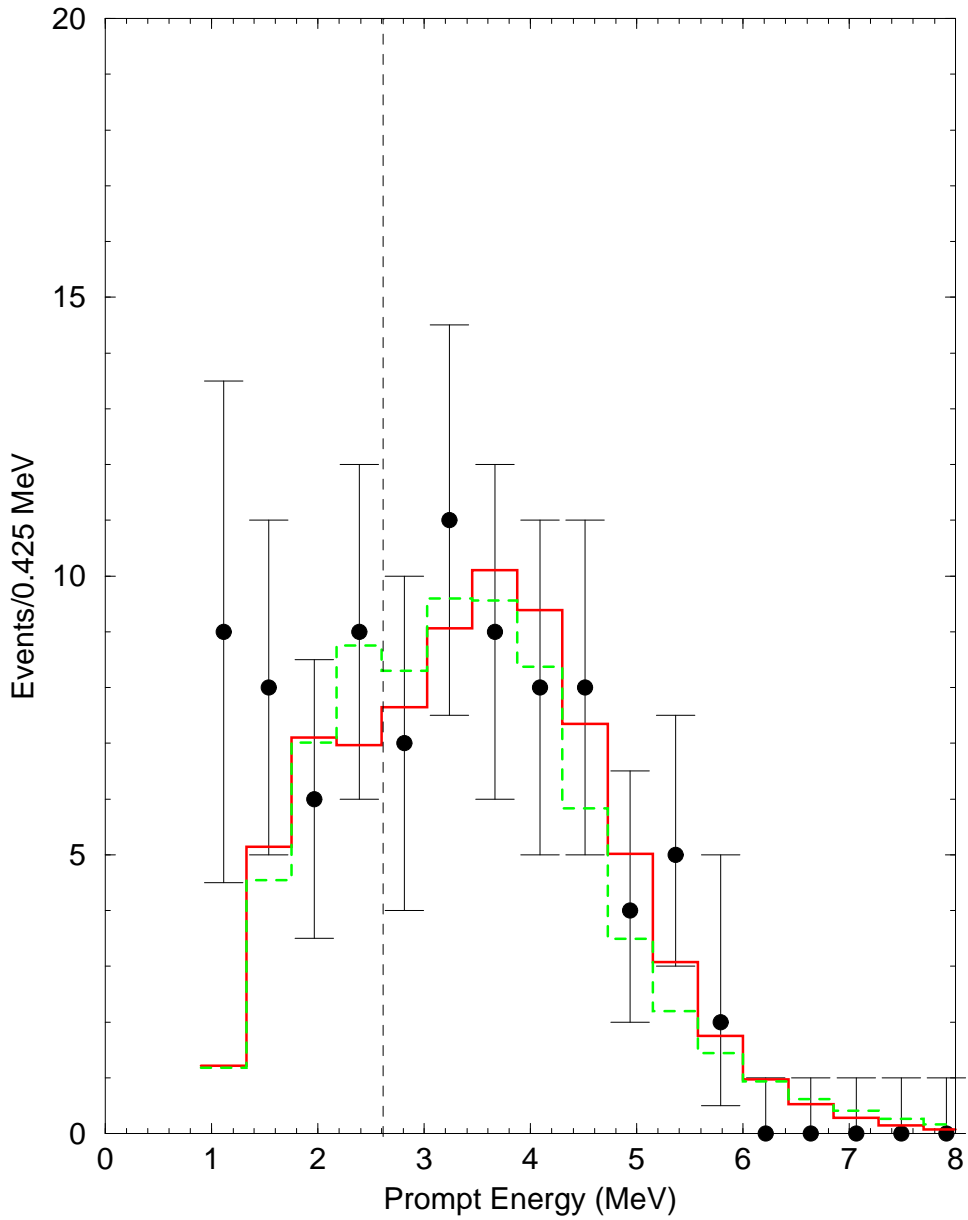}
\caption{The expected prompt energy spectra  
for the best fit points from the l-region (solid histogram) and
h-region (dashed histogram).  Also shown are the KamLAND experimental
points.  }
\label{fig:spectrum}
\end{figure}

%%%%%%%%%%%%%%%%%%%%%%%%%%%%%%%%%%%%%%%%%%%%%%%%%%%
\noindent
{\it Features of spectrum distortion}. 
In the fig.~\ref{fig:spectrum} we show the prompt energy spectra of
events for the best fit  points  in the l- and h-regions,  $N_l(E_p)$
and $N_h(E_p)$. The  spectra can be well understood in
terms of the effective oscillation phase $\phi_{eff}$: 
\be
N(E_p) \sim  N_0(E_p) \left[1 + D(E_p) \sin^2 \phi_{eff} \right],  
\label{specapp}
\ee 
where $D(E_p)$ is the averaging factor. 

In the best fit point of l-region (\ref{hth4}),  the peak at $E_p
\approx 3.6$ MeV corresponds to the oscillation maximum $\phi_{eff} =
\pi$ (exact position of maximum of the survival probability 
is at $E_p = 4.3 {\rm MeV}$). 
The closest oscillation minimum (phase $\phi_{eff} = 3\pi/2$) is
at $E_p \approx 2.4$ MeV and the next maximum ($\phi_{eff} = 2\pi$) is
at $E_p \approx 1.8$ MeV. Due to strong averaging effect the
structures below (in $E_p$) the main maximum  are not profound and
look more like a shoulder below the peak.  The first oscillation minimum 
is at $E_p = 7.2$ MeV.

In the case of h-region (\ref{hth5}) the spectrum has similar
structure but (due to larger $\Delta m^2$) the energy intervals
between maxima and minima decrease.  The main peak at $E_p \approx
3.4$ MeV corresponds now to the second oscillation maximum with
$\phi_{eff} = 2\pi$.  One can calculate then that the next minimum (
$\phi_{eff} = 5\pi/2$) is at $E_p \approx 2.7$ MeV and the next
maximum ($\phi_{eff} = 3\pi$) is at $E = 2.2$ MeV.  The higher energy
oscillation minimum ($\phi_{eff} = 3\pi/2$) is at $E = 5$ MeV.  This
pattern can be seen in the fig.~\ref{fig:spectrum} which confirms
validity of the effective phase consideration.

The measured spectrum, indeed,  gives a hint of existence of the low
energy shoulder.  Evidently with the present data it is impossible to
disentangle the l- and h- spectra.  Substantial decrease of errors is
needed.  Also decrease of the energy threshold will help.  According
to fig.~\ref{fig:spectrum}, $N_l(E_p) > N_h(E_p)$ at $E_p > 3.5$ MeV,
and $N_l(E_p) < N_h(E_p)$ at lower energies, especially in the interval
$E_p = (2.0 - 2.5)$ MeV.  Therefore  for the low threshold the difference
between $N_l(E_p)$ and $ N_h(E_p)$ can not be eliminated by
normalization (mixing angle). See similar discussion in~\cite{fogli}.\\

%%%%%%%%%%%%%%%%%%%%%%%%%%%%%%%%%%%%%%%%%%%%%%%%%%%%%%%%%%%%%%%%%
\noindent 
{\it Pull-off diagrams.} 
To check the quality of the fit we have constructed the 
pull-off diagrams~\cite{plamen} 
which show  deviations, $D_K$, of the predicted values of the observables 
$K_{bf}$ from the central experimental values, $K_{exp}$, in the units of 
the $1\sigma$ experimental errors, $\sigma_K$,: 
\be
D_K \equiv \frac{K_{bf}- K_{exp}}{\sigma_K}, 
\ee
where $K = Q_{Ar}, ~Q_{Ge},~ CC/NC, ~A_{DN}^{SNO},~A_{DN}^{SK},~R_{KL},~f_B$. 

\begin{figure}[ht]
\centering\leavevmode
\epsfxsize=.8\hsize
\epsfbox{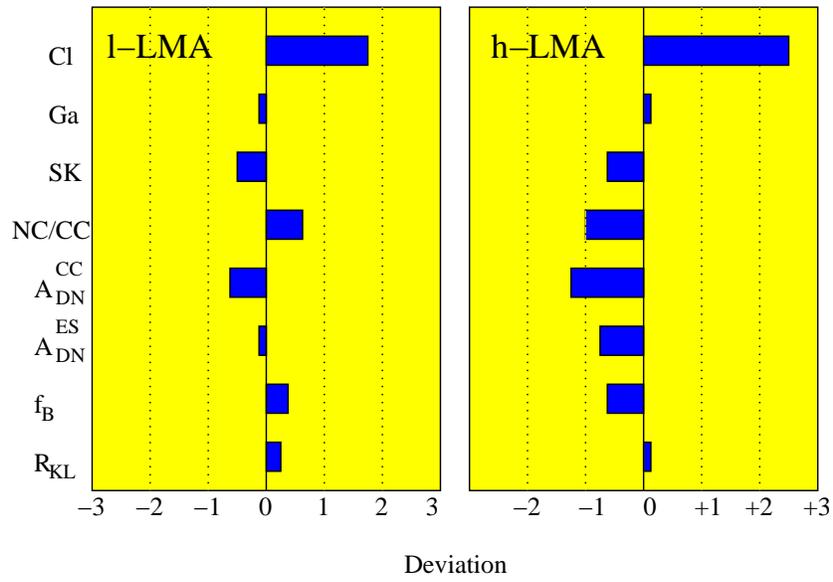}
\caption{The pull-off diagrams for the best fit points in the 
l- and h- regions. Shown are  deviations of the predicted values 
of different observables from the central experimental values 
in the units of $1\sigma$ (experimental).}
\label{pull}
\end{figure}

In fig.~\ref{pull} we show the pull-off diagrams for the best fit 
points from the l- and h-regions.  In both regions the largest deviation is 
for the $Ar$-production rate. In the h-region:  $Q_{Ar} > 3.1$
SNU, which is $2.5\sigma$ above the Homestake result.  
In the l-region the $Q_{Ar}$ deviation is smaller: about $1.8\sigma$. 

Other deviations are about $1\sigma$ or smaller;  
they are even smaller in  the  l-region. 
In the h-region a very small D-N asymmetry is expected. 
The l- and h-regions lead to the opposite sign deviations
for the CC/NC ratio at SNO, and therefore future precise measurements 
of this ratio will discriminate among l- and h- solutions. 
Notice also that in  h-region $f_B < 1$.

%%%%%%%%%%%%%%%%%%%%%%%%%%%%%%%%%%%%%%%%%%%%%%%%%%%%%%%%%%%%%%%%%%%%%%%%%
\section{Effect of 1-3 mixing}
%%%%%%%%%%%%%%%%%%%%%%%%%%%%%%%%%%%%%%%%%%%%%%%%%%%%%%%%%%%%%%%%%%%%%%%%%

We assume that $\Delta m^2_{13} = \Delta m^2_{atm} = (2 - 3) \cdot
10^{-3}$ eV$^2$, and there is a non-zero admixture of $\nu_e$ in the
third mass eigenstate described by the angle $\theta_{13}$.  Both for
KamLAND and for solar neutrinos the oscillations driven by $\Delta
m^2_{13}$ are averaged out and signals are determined by the survival
probabilities
\be
P_{ee} = (1 - \sin^2 \theta_{13})^2 {P}_{2} + \sin^4 \theta_{13}
\approx (1 - 2\sin^2 \theta_{13}) {P}_{2}.
\label{threepr}
\ee
Here $P_{2} = P_{2}(\Delta m^2_{12}, \theta_{12})$ is the two neutrino
vacuum oscillation probability for KamLAND and 
it is the two neutrino conversion probability for solar neutrinos.  The factor $(1 - 
\sin^2 \theta_{13})^2$ leads to additional suppression of the KamLAND signal
and shift of the allowed regions to smaller $\theta_{12}$ (see
detailed study in~\cite{ue3}).  The effect of $\theta_{13}$ on the 
solar neutrino analysis has been discussed in \cite{us:msw}.

We have performed the combined analysis of the solar neutrino and
KamLAND data for a fixed value of $\sin^2 \theta_{13}$ taken at the
upper bound given by the CHOOZ experiment~\cite{chooz}: $\sin^2
\theta_{13} = 0.04$. The number of degrees of freedom is the same as
in the $2\nu$ fit and we follow procedure described in
sect. \ref{solakam} with survival probabilities modified according to
(\ref{threepr}).

\begin{figure}[ht]
\centering\leavevmode
\epsfxsize=.8\hsize
\epsfbox{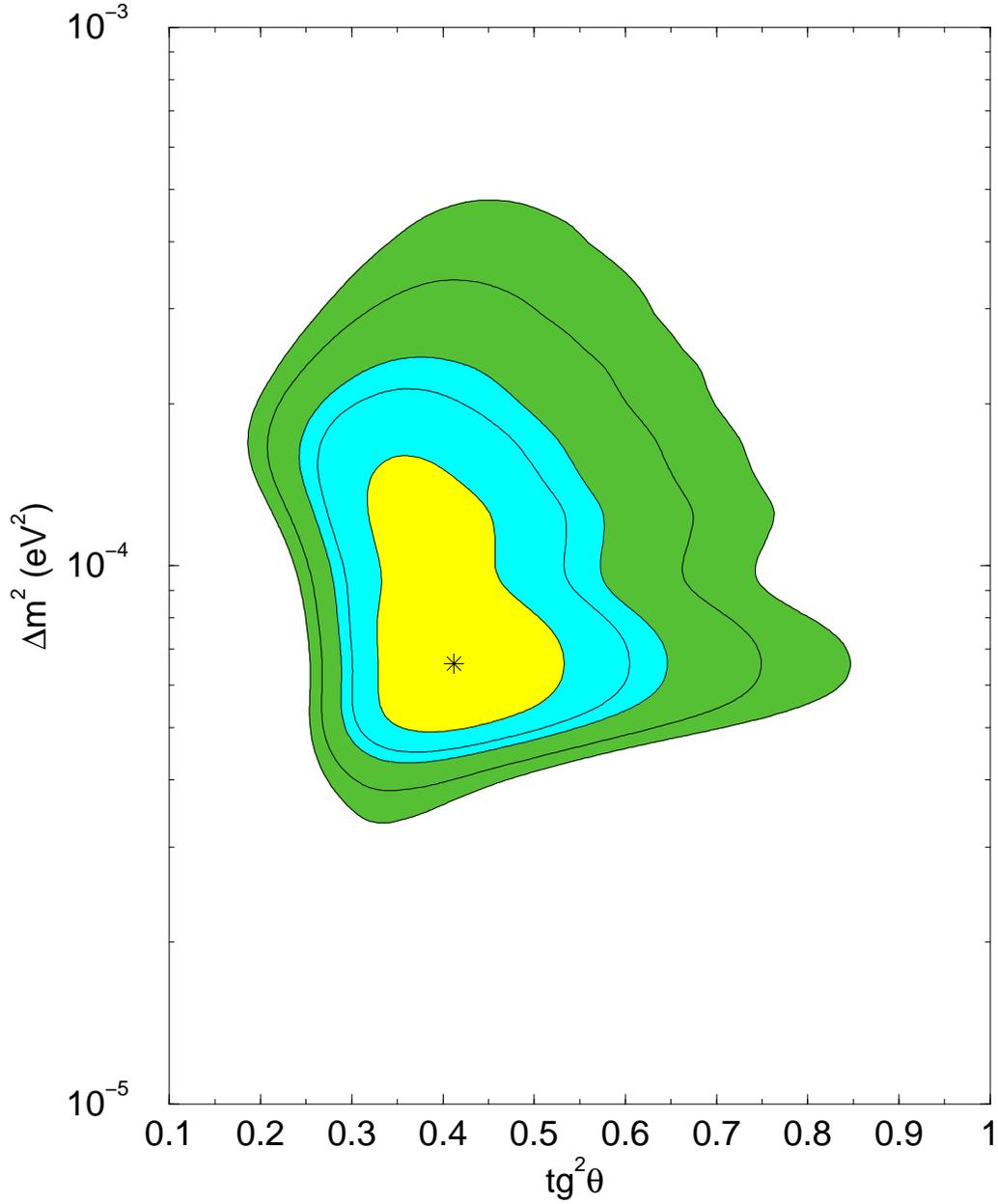}
\caption{Three neutrino analysis with $\sin^2\theta_{13}=0.04$. 
The allowed regions in $\tan^2\theta - \Delta m^2$ from a combined fit
of the solar neutrino data and the KamLAND rate at the 1$\sigma$,
90\%, 95\%, 99\% and 3$\sigma$ C.L..  }
\label{fig:chi2combrates_ue3}
\end{figure}

In fig.~\ref{fig:chi2combrates_ue3} we show results of the combined
analysis of the solar data and the KamLAND rate. As can be concluded
from comparison of fig.~\ref{fig:chi2combrates} and
fig.~\ref{fig:chi2combrates_ue3},  the main impact of non-zero 1-3 mixing
is a slight shift of the allowed region to larger $\Delta m^2$ and
smaller $\tan^2 \theta$. In particular, we find that the best fit, 
$\chi^2_{min}/d.o.f. = 66.21/79$, is at
\be 
\Delta m^2 =  6.60 \cdot 10^{-5} {\rm eV}^2 , 
~~~\tan^2 \theta = 0.411~~~ f_B = 1.08.
\label{bfp_ue3} 
\ee 

\begin{figure}[ht]
\centering\leavevmode
\epsfxsize=.8\hsize
\epsfbox{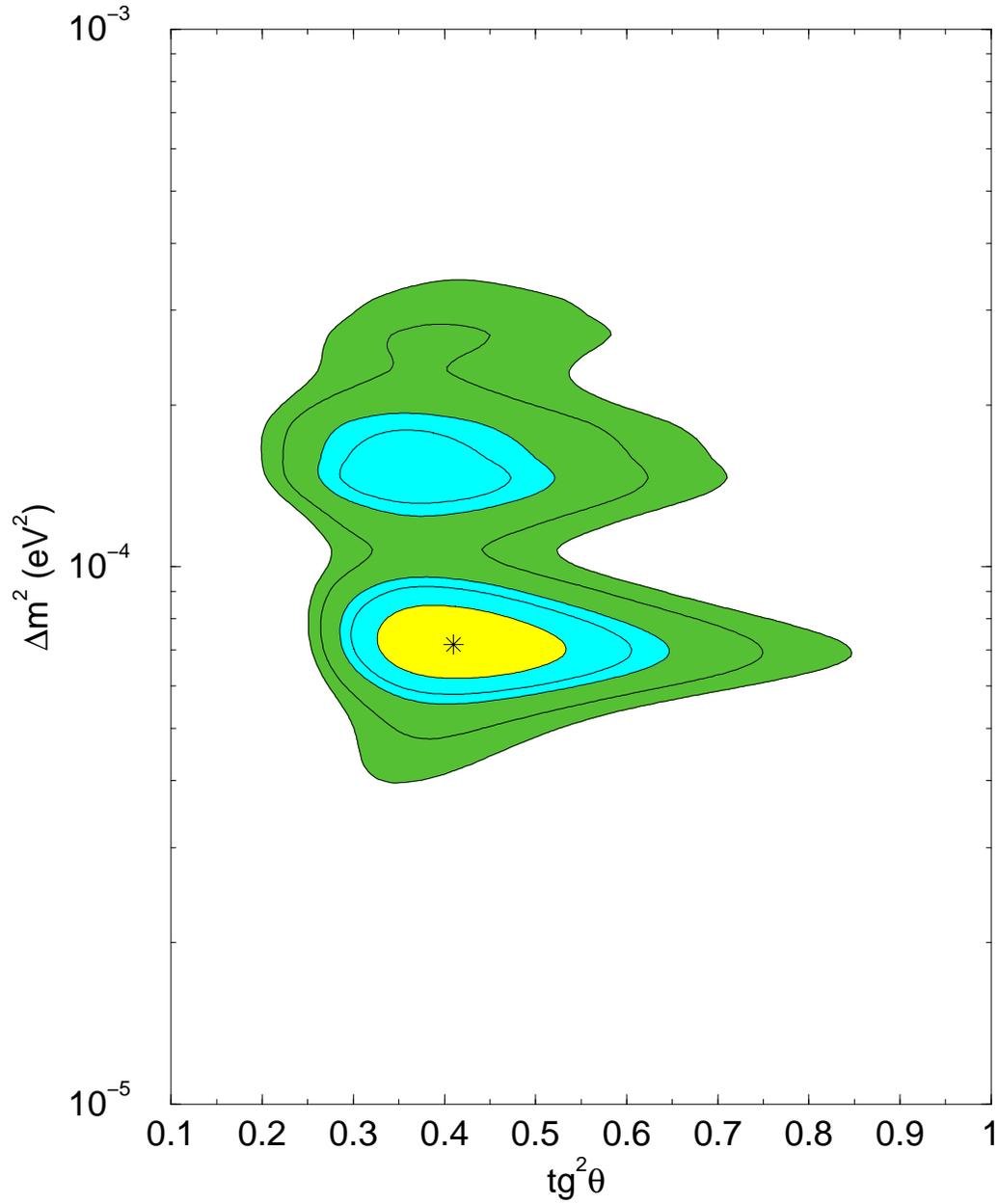}
\caption{
Three neutrino analysis with $\sin^2\theta_{13}=0.04$.  The allowed regions
in $\tan^2\theta - \Delta m^2$ plane from a combined analysis of the
solar neutrino data and the KamLAND spectrum at the 1$\sigma$, 90\%, 95\%,
99\% and 3$\sigma$ C.L..  }
\label{fig:chi2combspec_ue3}
\end{figure}

In fig.~\ref{fig:chi2combspec_ue3} we show result of the combined
analysis of the solar neutrino data and the KamLAND spectrum. From
fig.~\ref{fig:chi2combspec_ue3} and fig.~\ref{fig:chi2combspec} we
observe the same effect: 1-3 mixing leads to a shift of the allowed
region to larger $\Delta m^2$ and smaller $\tan^2 \theta$.  
In the best fit point we find  $\chi^2_{min}/d.o.f. = 69.24/91$ and
\be
\Delta m^2 =  7.17 \cdot 10^{-5}  {\rm eV}^2 , 
~~~ \tan^2 \theta = 0.41, ~~f_B = 1.07.   
\label{3spel}
\ee  

The introduction of the 1-3 mixing slightly worsen the
fit: $\Delta \chi^2 \approx 1$, it requires higher original boron
neutrino flux, than in the $2\nu$- case. At the same time, the 1-3 mixing 
improves a fit  in the h-region: It is accepted now at 90 \%
C.L. with respect to the best fit point (\ref{3spel}). This region
merges with the small spot at high $\Delta m^2$. The best fit point in the
h-region is shifted to smaller mixing: $\tan^2 \theta = 0.36$.

For convenience, the results of different fits are summarized in the
Table 1.  It shows high stability of the extracted parameters with
respect to a type of analysis.

\begin{table}
\begin{center}
\begin{tabular}{lcccc}
\hline 
Type of data fit & $\Delta m^2$ & $\tan^2\theta$ & $f_B$ &
$\chi^2_{min}$/d.o.f.  \\
\hline 
solar 2$\nu$ & 6.15 & 0.406 & 1.05 & 65.2/78 \\
\hline 
KL spectrum, 2$\nu$ & 7.34 & 0.453 & - & 2.92/11 \\
\hline 
solar + KL rate, 2$\nu$ & 6.03 & 0.411 & 1.05 & 65.3/79 \\
\hline 
solar + KL spectrum, 2$\nu$ & 7.32 & 0.409 & 1.05 & 68.2/91 \\
\hline
solar + KL rate, 3$\nu$ & 6.60 & 0.411 & 1.08 & 66.2/79 \\
\hline 
solar + KL spectrum, 3$\nu$ & 7.17 & 0.409 & 1.07 & 69.2/91 \\
\hline
\end{tabular}
\label{tab:chi2}
\caption{The parameters of the best fit points as well as $\chi^2_{min}/d.o.f.$
from different analyses of the data; $\Delta m^2$ is in the units
$10^{-5}$ eV$^2$.}
\end{center}
\end{table}
%%%

%%%%%%%%%%%%%%%%%%%%%%%%%%%%%%%%%%%%%%%%%%%%%%%%%%%%%%%%%%%%%%%%%%
\section{Next step}
%%%%%%%%%%%%%%%%%%%%%%%%%%%%%%%%%%%%%%%%%%%%%%%%%%%%%%%%%%%%%%%%%%

The key problems left after the first KamLAND results are

\begin{itemize}

\item 
more precise determination of the neutrino parameters: in particular,
(i) precise determination of the deviation of 12-mixing from maximal
mixing, (ii) strengthening of the upper bound on $\Delta m^2$ (iii)
discrimination between the two existing regions;
% determination of the best fit point. 

\item 
searches for effects beyond the single $\Delta m^2$ and single mixing
approximation;

\item
searches for differences of the neutrino oscillation parameters determined from
KamLAND and from the solar neutrino experiments.

\end{itemize}

Notice that precise knowledge of the  parameters is crucial not only for
the neutrinoless double beta decay searches, long baseline
experiments, studies of the atmospheric and supernova 
neutrinos, etc., but also for understanding of physics of the solar
neutrino conversion.  In the region of the best fit point a dominating
process (at least for $E > (0.5 - 1)$ MeV) is the adiabatic neutrino
conversion (MSW), whereas in the high $\Delta m^2$ region allowed by
KamLAND at the $2\sigma$ level, the effect is reduced 
to the averaged vacuum oscillations (a la Gribov-Pontecorvo)~\cite{pontecorvo} 
with small matter corrections.

In this connection, we will discuss two questions.\\

%%%%%%%%%%%%%%%%%%%%%%%%%%%%%%%%%%%%%%%%%%%%%%%%%%%%%%%%%%%%
\noindent
{\it How small $\Delta m^2$ can be?}  This is especially important
question, {\it e.g.}, for measurements of the earth regeneration
effect.  In the l-region we get
\be
\Delta m^2 > 4 \cdot 10^{-5}~~ {\rm eV}^2, ~~~  (3\sigma)  
\label{bfp2}
\ee
and it is difficult to expect that lower values will be allowed.  

The bound (\ref{bfp2}) appears as an interplay of both the 
KamLAND rate and the shape. As it follows from the fig.~\ref{fig:chi2combrates}, 
the KamLAND rate strengthens  the lower bounds: 
$\Delta m^2 \geq (3.0,~4.0,~4.7) \times 10^{-5}$ eV$^2$, 
at the $1\sigma$, $2\sigma$, $3\sigma$ correspondingly 
which should be compared with 
$\Delta m^2 \geq (2.5,~3.2,~4.0) \times 10^{-5}$ eV$^2$ 
from the solar analysis only.   
Adding the spectral data results in the bounds 
$\Delta m^2 \geq 3.3,~~5.5,~~6.2 ~~ \times 10^{-5}$ eV$^2$  

With decrease of $\Delta m^2$ the oscillatory pattern of the spectrum
shifts to lower energies.  For $\Delta m^2 = 5 \cdot 10^{-5}$ eV$^2$
the maximum of spectrum is at $E_p = 2.7$ MeV and the oscillation
suppression increases with energy \cite{us:msw}. The oscillation
minimum is at $E_p \approx 5$ MeV.  If $R_{KL} (2.7~ {\rm MeV}) =
0.81$, then $R_{KL} (4.0 ~{\rm MeV}) = 0.47$. The KamLAND spectrum does
not show such a fast decrease.

One can characterize the  spectrum distortion 
by a relative suppression of signal at the high  (say, above  4.3 
MeV) and at the low (below 4.3 MeV) energies
~\footnote{An alternative way is to introduce the moments of the 
energy spectrum~\cite{john2}.}. 
The energy interval (2.6 - 4.3) MeV contains KamLAND energy 4 bins.   
Introducing the suppression 
factors $R_{KL} (< 4.3~{\rm MeV})$ and  $R_{KL} (> 4.3~{\rm MeV})$ 
we can define the ratio
\be
k = \frac{1 - R_{KL} (> 4.3~{\rm MeV})}{1 - R_{KL} (< 4.3~{\rm MeV})}.
\ee
$k$ which we will call the {\it shape parameter} 
does not depend on the mixing angle and normalization of spectrum. 
It increases with increase of the oscillation suppression  
at high energies. 

Using the KamLAND data we get the experimental value
\be 
k^{exp} = 0.84^{+.42}_{-0.35}, ~~~1\sigma.
\ee

\begin{figure}[ht]
\centering\leavevmode
\epsfxsize=.8\hsize
\epsfbox{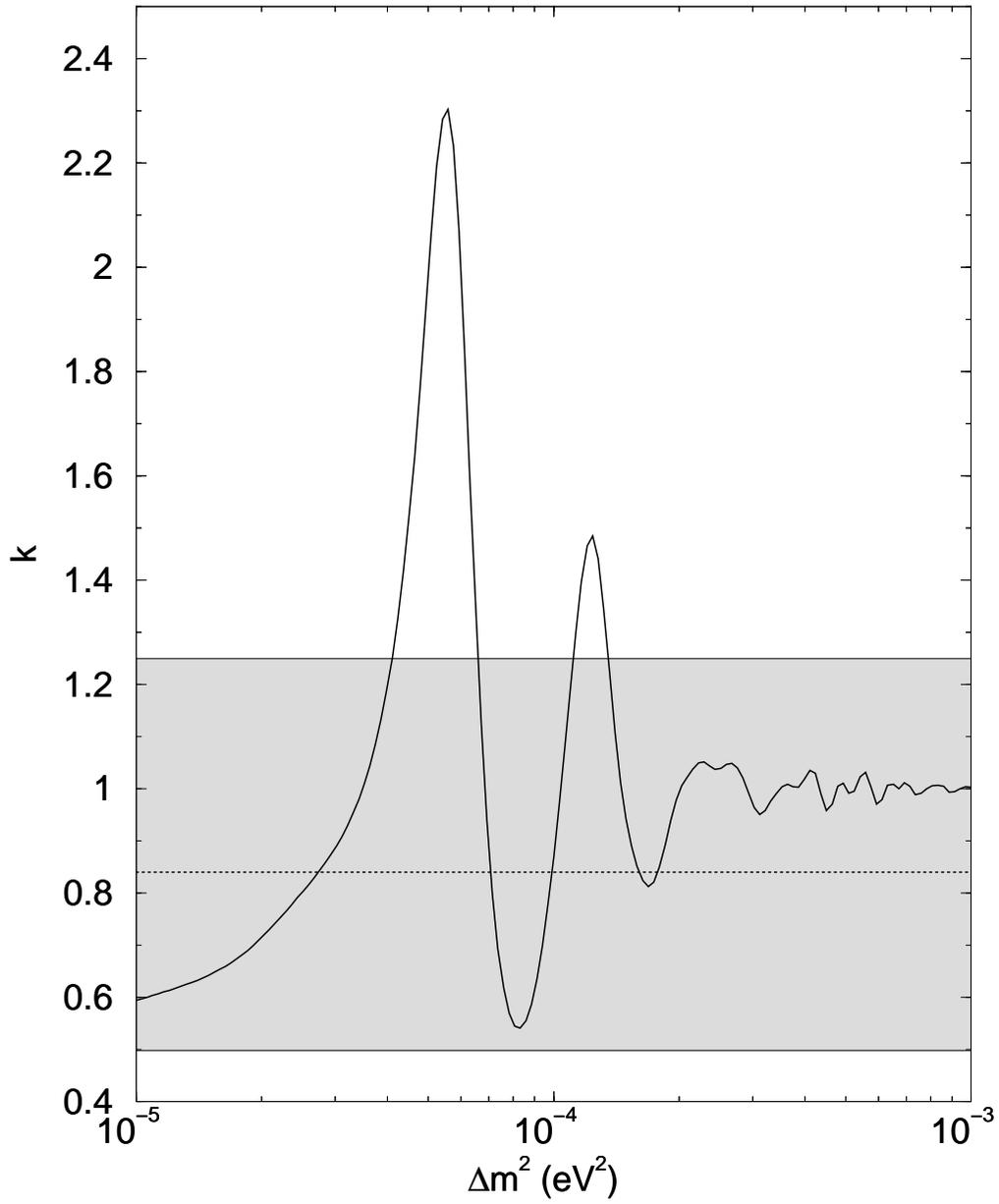}
\caption{The dependence of
the shape parameter $k$ on  $\Delta m^2$ for 
 $\tan^2\theta=0.41$. Shown are the central experimental value (dotted
 line) and the $1\sigma$ experimental band (shadowed).
}
\label{fig:k}
\end{figure}

In fig.~\ref{fig:k} we present the  dependence of the shape parameter  
$k$ on  $\Delta m^2$ for fixed mixing: $\tan^2 \theta = 0.41$. 
For the spectrum which corresponds to the best combined fit we find  $k =
0.70$, whereas for $\Delta m^2 = 5\cdot 10^{-5}$ eV$^2$ the ratio equals $k = 2.0$. 
For the h-region best point: $k = 0.94$. 

Notice that in the l-region below $\Delta m^2 = 8 \cdot 10^{-5}$ eV$^2$, 
$k$ increases quickly with decrease of $\Delta m^2$, reaching a maximal value at 
$\Delta m^2 = 5.5 \cdot 10^{-5}$ eV$^2$. Below that, 
the parameter $k$ decreases with $\Delta m^2$. In this region, however, 
the total event  rate  decreases fast giving the bound on $\Delta m^2$. 
This explains a shift of the allowed (at $3\sigma$)  region to smaller 
mixings with decrease of $\Delta m^2$.  

Notice also that the central experimental value of $k$ can be reproduced 
in the both allowed regions (l- and h-). Therefore future precise measurement 
of spectrum will further sharpen determination of $\Delta m^2$ 
within a given island. To discriminate among the islands one needs to 
use more elaborated criteria (not just $k$) or a complete spectral information.\\ 

%%%%%%%%%%%%%%%%%%%%%%%%%%%%%%%%%%%%%%%%%%%%%%%%%%%%%%%%%%%%%%%%%
\noindent
{\it How large is the large mixing?}  In contrast to \cite{kamf} our best
fit point is at non-maximal mixing $(\sin^2 2\theta = 0.86)$ being
rather close to the best fit point from the solar neutrino analysis. 
Similar deviation from maximal mixing has been obtained in our
rate+spectrum analysis.  Notice that the KamLAND data have weak
sensitivity to the mixing (weaker than the solar neutrino data).  The
allowed regions cover the interval
\be
\tan^2 \theta = 0.12 - 1.00 ~~~~~(\theta < \pi/2), ~~~~ 95 \% {\rm C.L.}. 
\ee
Even at $1\sigma$ the interval $\tan^2 \theta = 0.23 - 1.00$ $(\theta
< \pi/2)$ is allowed.  The reason is that KamLAND is essentially
the  vacuum oscillation experiment (see evaluation of matter effects in
KamLAND in \cite{BGP}), and  effects of the vacuum oscillations depend
on deviation from maximal mixing which can be characterized by
$\epsilon = (1/2 - \sin^2\theta)$ quadratically: $P \propto 1 -
4\epsilon^2$.  The matter conversion depends on $\epsilon$ linearly:
$P \propto 1 - 2\epsilon$ \cite{nir}.

\begin{figure}[ht]
\centering\leavevmode
\epsfxsize=.8\hsize
\epsfbox{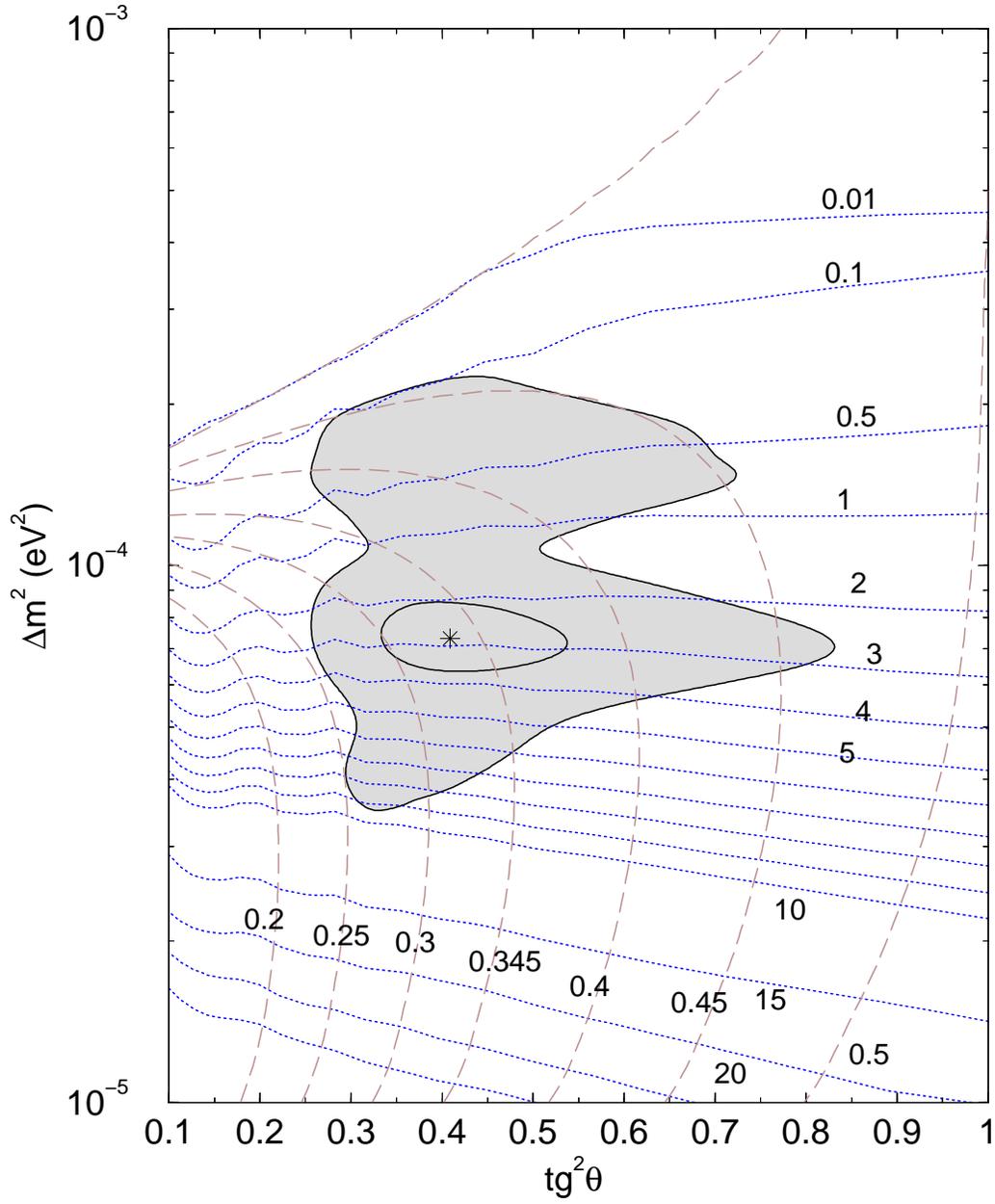}
\caption{Predictions for the CC/NC ratio and the Day-Night asymmetry at 
SNO. The dashed lines are the lines of constant CC/NC ratio (numbers
at the curves) and the dotted lines show the lines of constant
$A_{DN}^{SNO}$ (numbers at the curves in \%).  We show also the
($1\sigma$ and $3\sigma$) allowed regions of the oscillation parameters
from the combined fit of the solar neutrino data and the KamLAND
spectrum. The best fit  point is indicated by a star.}
\label{fig:previsions}
\end{figure}

Notice that maximal mixing is rather strongly disfavored by all measured solar
neutrino rates.  In the point indicated by KamLAND we predict the
charged current to neutral current measurement at SNO, the Argon
production rate and the Germanium production rate:
\be
\frac{\rm CC}{\rm NC} = 0.51~(+3.3 \sigma) ,~~~  
Q_{Ar} = 3.2~{\rm  SNU} ~(+2.5 \sigma), ~~~Q_{Ge} = 63~{\rm SNU}~(-1.6 \sigma).
\ee
In brackets we show the pulls of the predictions from the best fit
experimental values. 
As follows from the fig.~\ref{fig:previsions} future precise measurements of the 
CC/NC ratio at SNO will strengthen the upper bounds on 
both mixing and $\Delta m^2$. \\

It is important to ``overdetermine" the neutrino parameters measuring all
possible observables.  This will allow us to make cross-checks of
selected solution and to search for  inconsistencies which
will require extensions of the theoretical context.  In this connection
let us consider predictions for the forthcoming measurements.

\noindent
{\it 1). Precise measurements of the CC/NC ratio at SNO.}  In
fig.~\ref{fig:previsions} we show the contours of constant CC/NC
ratio.  We find predictions for the best fit point and the $3\sigma$
interval:
\be
\frac{\rm CC}{\rm NC} = 0.33^{+ 0.15}_{-0.07} ~, ~~~ 3\sigma . 
\ee
At $99\%$ C.L. the islands split and we get the following predictions
for them separately:
\be
\frac{\rm CC}{\rm NC}({\rm l-region}) = 0.33^{+ 0.11}_{-0.06}~, ~~~~ 
\frac{\rm CC}{\rm NC}({\rm h-region}) = 0.38^{+ 0.06}_{-0.03}~, ~~~~~~~~ 
99\% ~{\rm C.L.} ~.
\ee
Values of CC/NC $< 0.35$ will exclude the h-region.  Precise
measurements of the ratio CC/NC will also strengthen the upper bounds
on mixing and $\Delta m^2$. \\

\noindent
{\it 2). The day-night asymmetry  at SNO.}  The
KamLAND provides a  strong lover bound on $\Delta m^2$, and shifts the
best fit point to larger $\Delta m^2$.  This further diminishes the
expected value of  day-night asymmetry.  In
fig.~\ref{fig:previsions} we show the contours of constant
$A_{DN}^{SNO}$.  The best fit point prediction and the $3\sigma$ bound
equal
\be
A_{ND}^{SNO} = 2.8 \pm 0.8 \% ~,~~~ (1 \sigma),~~~~~ A_{ND}^{SNO} < 9
\% ~~~~ (3 \sigma).
\label{dnas}
\ee
The present best fit value of the SNO asymmetry, $ 7\%$, is accepted
at about 99\% C.L..  Observations of the asymmetry $A_{ND}^{SNO} > 1
\%$ will exclude the h-region.  The expected asymmetry at
Super-Kamiokande is even smaller: In the best fit point we expect
$A_{ND}^{SK} \approx (1.7 - 2.0) \%$.\\

\noindent
{\it 3). The turn up of the spectrum at SNO and Super-Kamiokande at
low energies.}  Using results of \cite{plamen} we predict for the best
fit point (\ref{hth4}) an increase of ratio of the observed to
expected (without oscillation) number of the CC events, $R_{SNO}^{CC}$, from 0.31 at 8 MeV
to 0.345 at 5 MeV, so that
\be
\frac{R_{SNO}^{CC}({\rm 5 MeV}) - R_{SNO}^{CC}({\rm 8 MeV})}
{R_{SNO}^{CC}({\rm 8 MeV})} = 0.10 - 0.12.
\ee
In Super-Kamiokande the turn up 
is about $(5 - 7)\%$ in the same interval (5 - 8) MeV.\\

\noindent
{\it 4). Further KamLAND measurements.}  Possible impact can be
estimated using figures \ref{fig:chi2kaml}, \ref{fig:chi2kaml_lt},
\ref{fig:spectrum}.  A decrease of the error by factor of 2 (which
will require both significant increase of statistics and decrease of
the systematic error) will allow KamLAND alone to exclude all the
regions but the l-region at 95 \% C.L., if the best fit is at the same point as it is 
determined now. \\

\noindent
{\it 5). BOREXINO.} At $3\sigma$ we predict the following suppressions
of signals with respect to the SSM predictions to BOREXINO
experiment~\cite{borexino}, in the two allowed regions:
\be
R_B({\rm l-region}) = 0.61 - 0.73, ~~~ R_B({\rm h-region}) = 0.62 -
0.73.
\ee
So, BOREXINO will perform consistency check but it will not
distinguish the l- and h- regions. \\

\noindent
{\it 6). Gallium production rate. }  In the best fit point one
predicts the germanium production rate $Q_{Ge} = 71$ SNU.  In the
h-region best fit point $Q_{Ge} = 72$ SNU.  So, the Gallium
experiments do not discriminate among the l - and h - regions.
However, precise measurements of $Q_{Ge}$ are important for
improvements of the bound on the 1-2 mixing and its deviation from maximal
value.

%%%%%%%%%%%%%%%%%%%%%%%%%%%%%%%%%%%%%%%%%%%%%%%%%%%%%%%%%%%%%%%%%%%%%%%%%%%
\section{Conclusions}
%%%%%%%%%%%%%%%%%%%%%%%%%%%%%%%%%%%%%%%%%%%%%%%%%%%%%%%%%%%%%%%%%%%%%%%%%%%

\noindent
1. The first KamLAND results (rate and spectrum) are in a very good
agreement with the predictions based on the LMA MSW solution of the
solar neutrino problem.\\

\noindent
2. Our analysis of the KamLAND data reproduces well the results of the
collaboration.  The oscillation parameters extracted from the KamLAND
data and from the solar neutrino data agree within $1\sigma$.\\

\noindent
3. We have performed a combined analysis of the solar and KamLAND
results.  The main impact of the KamLAND results on the LMA solution 
can be summarized in the following way. KamLAND

\begin{itemize}

\item
shifts  the $\Delta m^2$ to slightly higher values $6.15 \cdot
10^{-5}$ $\rightarrow 7.3 \cdot 10^{-5}$ eV$^2$; (the mixing is
practically unchanged: $\tan^2 \theta = 0.41$, and this number is
rather stable with respect to variations of the analysis;

\item
establishes rather solid lower bound on the $\Delta m^2$: $\Delta m^2
> 4 \cdot 10^{-5}$ eV$^2$ ($3\sigma$);

\item 
disintegrates  the LMA region at 99\% C.L.  into two parts.

\end{itemize}

KamLAND further disfavors high values of $\Delta m^2$: $\Delta m^2 >
3\cdot 10^{-4}$ eV$^2$.\\

\noindent
4. Inclusion of the 1-3 mixing   shifts the allowed regions to
larger $\Delta m^2$ and smaller $\tan^2 \theta$. $\chi^2$ slightly
increases with $\sin^2 \theta_{13}$.\\

\noindent
5. The KamLAND results strengthen the upper bound on the expected
value of the day-night asymmetry at SNO: $A_{DN}^{SNO} < 9 \%$.  We
predict about 10\% turn up of the energy spectrum at SNO in the
interval of energies (5 - 8) MeV.  The CC/NC ratio is expected to be
CC/NC $\approx 0.33$, that is, near the present best SNO value.

Future SNO  measurements of the CC/NC ratio and 
$A_{DN}^{SNO}$ will have further strong impact 
on the LMA parameter space.

\section{Acknowledgments}

The authors are grateful to E. Kh. Akhmedov for useful discussions.

%%%%%%%%%%%%%%%%%%%%%%%%%%%%%%%%fff1%%%%%%%%%%%%%%%%
\newpage

%%%%%%%%%%%%%%%%%%%%%%%%%%%%%%%%fff2
\newpage

%%%%%%%%%%%%%%%%%%%%%%%%%%%%%%%fff3
\newpage

%%%%%%%%%%%%%%%%%%%%%%%%%%%%%%%fff4
\newpage

%%%%%%%%%%%%%%%%%%%%%%%%%%%%%%%fff5
\newpage

%%%%%%%%%%%%%%%%%%%%%%%%%%%%%%%%%%%%%%%%%%%%%%%%%%%%%%%%%%%%%%%%%%%%%%%%%%%%%%%%%
\newpage

%%%%%%%%%%%%%%%%%%%%%%%%%%%%%%%%%%%%%%%%%%%%%%%%%%%%%%%%%%%%%%%%%%%%%%%%%%%%%%%%%%%%%

%%%%%%%%%%%%%%%%%%%%%%%%%%%%%%%fff6
\newpage

%%%%%%%%%%%%%%%%%%%%%%%%%%%%%%%fff7
\newpage

%%%%%%%%%%%%%%%%%%%%%%%%%%%%%%%fff8
\newpage

%%%%%%%%%%%%%%%%%%%%%%%%%%%%%%%fff9
\newpage

%%%%%%%%%%%%%references%%%%%%%%%%%%%%%%%%%%%%%%%%%%%%%%%%%%%%%%%%

\end{document}